\documentclass[12pt]{iopart}
\usepackage{graphicx,cite}
\begin{document}

\title[1D hydrogen atom with minimal length uncertainty and maximal momentum]
{One-dimensional hydrogen atom with minimal length uncertainty and
maximal momentum}

\author{Pouria Pedram}
\address{Department of Physics, Science and Research Branch, Islamic Azad University, Tehran, Iran} \ead{p.pedram@srbiau.ac.ir}

\begin{abstract}
We present exact energy eigenvalues and eigenfunctions of the
one-dimensional hydrogen atom in the framework of the Generalized
(Gravitational) Uncertainty Principle (GUP). This form of GUP is
consistent with various theories of quantum gravity such as string
theory, loop quantum gravity, black-hole physics, and doubly special
relativity and implies a minimal length uncertainty and a maximal
momentum. We show that the quantized energy spectrum exactly agrees
with the semiclassical results.
\end{abstract}

\pacs{03.65.Ge, 03.65.Fd, 03.65.Sq} \vspace{2pc}

\section{Introduction}
In recent years, the investigation of the effects of the Generalized
Uncertainty Principle (GUP) on various physical systems has
attracted much attention and many authors have found exact or
approximate solutions in both classical and quantum mechanical
domains \cite{felder,epl1,epl2,jhep}. Indeed, because of the
universality of this gravitational effect, it couples to all forms
of matter and modifies the corresponding Hamiltonians in both
non-relativistic and relativistic limits. Moreover, the existence of
a finite lower bound to the possible resolution of length
proportional to the Planck length
$\ell_{Pl}=\sqrt{\frac{G\hbar}{c^3}}\approx 10^{-35}$m, where $G$ is
Newton's gravitational constant, naturally arises from various
candidates of quantum gravity such as string theory
\cite{1,2,3-1,3-2,3-3,4}, loop quantum gravity \cite{5}, and
noncommutative spacetime \cite{6,7,8}. Also the presence of a
maximal momentum proportional to $\ell_{Pl}^{-1}$ is in agreement
with Doubly Special Relativity (DSR) theories \cite{21,22,23}.

The problem of the hydrogen atom is studied in ordinary quantum
mechanics and its well-known exact energy eigenvalues and
eigenfunctions have already been obtained \cite{11,12,13,14,15}. In
the presence of the minimal length uncertainty, this problem is also
studied exactly and perturbatively in
Refs.~\cite{Akhoury,Bouaziz,Brau,Benczik,Tkachuk}. Moreover, using
the formally self-adjoint representation \cite{pedramPRD,pedramPLB},
the quantization condition is completely determined in one-dimension
upon imposing the Hermicity condition on the GUP-corrected
Hamiltonian \cite{pedramH}.

In this paper, we consider a recently proposed generalized
uncertainty principle that implies both a minimal length uncertainty
proportional to $\hbar\sqrt{\beta}$ and a maximal momentum
proportional to $\frac{1}{\sqrt{\beta}}$ where $\beta$ is the
deformation parameter \cite{pedramHigh,pedramHigh2}. The problems of
the free particle, particle in box, harmonic oscillator, maximally
localized states, black-body radiation, and cosmological constant
have been studied in this framework \cite{pedramHigh,pedramHigh2}.
Here, we solve the problem of the one-dimensional hydrogen atom in
this deformed quantum mechanics and find the exact energy
eigenvalues and eigenfunctions in the momentum space. We show that
imposing the Hermicity condition on the Hamiltonian results in the
self-adjointness of the Hamiltonian and the vanishing of the wave
functions at the origin in coordinate space. We finally obtain the
semiclassical energy spectrum that exactly agrees with the quantum
mechanical results as well as with ordinary quantum mechanics and
the modified quantum mechanics with just a minimal length.

\section{Momentum space representation}
Consider the following one-dimensional commutation relation (see
Refs.~\cite{pedramHigh,pedramHigh2} for details):
\begin{eqnarray}\label{gupc}
[X,P]=\frac{i\hbar}{1-\beta P^2},
\end{eqnarray}
which to the first order of the GUP parameter agrees with the
well-known GUP proposal by Kempf, Mangano and Mann \cite{7}. To
satisfy the above commutation relation, we write the position and
momentum operators in the momentum space representation as
\begin{eqnarray}\label{rep1}
P \phi(p)&=& p\,\phi(p),\\
X\phi(p)&=& \frac{i\hbar}{1 - \beta
p^2}\partial_p\phi(p).\label{rep2}
\end{eqnarray}
Now the position operator is symmetric subject to the following
scalar product
\begin{eqnarray}\label{comp2}
\langle\psi|\phi\rangle=\int_{-1/\sqrt{\beta}}^{+1/\sqrt{\beta}}\,\mathrm{d}
p \left(1-\beta p^2\right)\psi^{*}(p)\phi(p),
\end{eqnarray}
and we have
\begin{eqnarray}
\langle p|p'\rangle= \frac{\delta(p-p')}{1-\beta p^2}.
\end{eqnarray}

In the framework of this generalized uncertainty principle, the
absolutely smallest uncertainty in position is given by
\cite{pedramHigh}
\begin{eqnarray}\label{minl}
(\Delta X)_{min}=\frac{3\sqrt{3}}{4}\hbar\sqrt{\beta}.
\end{eqnarray}
and the maximal momentum is
\begin{eqnarray}
P_{max}=\frac{1}{\sqrt{\beta}}.
\end{eqnarray}

Now consider the one-dimensional hydrogen atom eigenvalue problem
\begin{eqnarray}\label{1}
P^2\phi-\frac{\alpha}{X}\phi=E\phi,
\end{eqnarray}
where we set $\hbar=1=2m$. In momentum space, the action of the
inverse operator $1/X$ is expressed as
\begin{eqnarray}\label{X}
\hspace{-1cm}\frac{1}{X}\phi(p)=-i\int_{-\frac{1}{\sqrt{\beta}}}^{p}\left(1-\beta
q^2\right)\phi(q)\,\mathrm{d}
q+c,\hspace{.3cm}\frac{-1}{\sqrt{\beta}}<p<\frac{+1}{\sqrt{\beta}},
\end{eqnarray}
where, as we shall show, $c$ is indeed a constant. The presence of
$c$ is due to the fact that the application of (\ref{X}) with $c=0$
only results in the trivial solution $\phi(p) = 0$ \cite{Tkachuk}.
Moreover, in the absence of GUP, this constant corresponds to the
derivative discontinuity of the eigenfunctions at the origin in the
coordinate representation \cite{12}. This definition implies
\begin{eqnarray}\label{x11}
X \frac{1}{X}\phi&=&\phi,\label{x12}\\
  \frac{1}{X}X\phi&=&\phi+c,\label{x13}\\
\left[X ,\frac{1}{X}\right]\phi&=&-c.
\end{eqnarray}
In the same way we have $X^\dagger\phi=\displaystyle\frac{i}{1-\beta
p^2} \frac{\partial \phi}{\partial p}$  and the action of the
adjoint of $1/X$ is given by
\begin{eqnarray}\label{X2}
\hspace{-1.5cm}\left(\frac{1}{X}\right)^\dagger
\phi(p)=-i\int_{-\frac{1}{\sqrt{\beta}}}^{p}\left(1-\beta
q^2\right)\phi(q)\,\mathrm{d}
q+c^*,\hspace{.3cm}\frac{-1}{\sqrt{\beta}}<p<\frac{+1}{\sqrt{\beta}},
\end{eqnarray}
Thus, we have
\begin{eqnarray}
X^\dagger\left(\frac{1}{X}\right)^\dagger\phi&=&\phi,\label{x21}\\
\left(\frac{1}{X}\right)^\dagger  X^\dagger\phi&=&\phi+c^*,\label{x22}\\
\left[X^\dagger ,\left(\frac{1}{X}\right)^\dagger
\right]\phi&=&-c^*\label{x23}.
\end{eqnarray}

At this point, we prove that $X^{-1}$ is not a linear operator. In a
basis which $X$  (a linear operator) is diagonal, the formal
operational relation $X \frac{1}{X}=1$ (\ref{x11}) implies that if
$X^{-1}$ is a linear operator with a matrix representation, it is
also diagonal. So we obtain
\begin{eqnarray}
[X,X^{-1}]=0,
\end{eqnarray}
which apparently contradicts Eq.~(\ref{x13}). The same argument also
applies for $X^\dagger$. Explicitly we have
\begin{eqnarray}
\fl\frac{1}{X}\left[\mu\phi(p)+\nu\varphi(p)\right]&=&-i\mu\int_{-\frac{1}{\sqrt{\beta}}}^{p}\left(1-\beta
q^2\right)\phi(q)\,\mathrm{d}
q-i\nu\int_{-\frac{1}{\sqrt{\beta}}}^{p}\left(1-\beta
q^2\right)\varphi(q)\,\mathrm{d}
q+c,\nonumber\\
&\ne&\mu\frac{1}{X}\phi(p)+\nu\frac{1}{X}\varphi(p),
\end{eqnarray}
and a similar relation for $\left(\frac{1}{X}\right)^\dagger$. Thus
$c$ is a constant not a linear operator, i.e., a linear functional.

The action of the position operator and its adjoint
(\ref{X},\ref{X2}) leads to
\begin{eqnarray}
\left[\frac{1}{X}-\left(\frac{1}{X}\right)^\dagger\right]\phi=2\mathrm{Im}[c].
\end{eqnarray}
Since the momentum operator $P$ is Hermitian, i.e. $P=P^\dagger$,
the requirement of the Hermicity for the Hamiltonian implies
\begin{eqnarray}\label{imc}
\mathrm{Im}[c]=0.
\end{eqnarray}
Here, as we shall see, by imposing the Hermicity condition, we
completely determines the quantization condition that uniquely
determines the energy spectrum.

The generalized Schr\"odinger equation in momentum space now reads
\begin{eqnarray}\label{Seq}
p^2\phi(p)+i\alpha\int_{-\frac{1}{\sqrt{\beta}}}^{p}\left(1-\beta
q^2\right)\phi(q)\,\mathrm{d} q-\alpha \,c=-\epsilon\,\phi(p),
\end{eqnarray}
where $\epsilon=-E$. If we differentiate this equation with respect
to $p$ we obtain
\begin{eqnarray}
\phi'(p)+\frac{2p+i\alpha\left(1-\beta
p^2\right)}{p^2+\epsilon}\phi(p)=0.
\end{eqnarray}
The solution then reads
\begin{eqnarray}\label{sol0}
\phi(p)=\frac{\mathcal{A}\,e^{i \alpha\beta p}}{p^2+\epsilon}
\exp\left[-i\frac{\alpha(1+\beta
\epsilon)}{\sqrt{\epsilon}}\arctan\left(\frac{p}{\sqrt{\epsilon}}\right)\right].
\end{eqnarray}
It can be expressed as
\begin{eqnarray}\label{sol}
\phi(p)=\frac{\mathcal{A}\,e^{i \alpha\beta
p}}{p^2+\epsilon}\left(\frac{1-i p/\sqrt{\epsilon}}{1+i
p/\sqrt{\epsilon}}\right)^{\frac{\alpha(1+\beta
\epsilon)}{2\sqrt{\epsilon}}},
\end{eqnarray}
where $\mathcal{A}$ is the normalization coefficient. Substituting
the above expression into the eigenvalue equation (\ref{Seq}) leads
to
\begin{eqnarray}
\hspace{-1cm}c=\frac{1}{\alpha}\,\,\lim_{p\rightarrow\frac{-1}{\sqrt{\beta}}}\left(p^2+\epsilon\right)\phi(p)=
\mathcal{A}\,e^{-i \alpha\sqrt{\beta}}
\left(\frac{\sqrt{\beta\epsilon}+i}{\sqrt{\beta\epsilon}-i}\right)^{\frac{\alpha(1+\beta
\epsilon)}{2\sqrt{\epsilon}}}.
\end{eqnarray}
Therefore, the probability density in momentum space is
\begin{eqnarray}\label{pd}
|\phi(p)|^2=\frac{\mathcal{A}^2}{\left(p^2+\epsilon\right)^2},
\end{eqnarray}
and the normalization coefficient can be written as
\begin{eqnarray}
\mathcal{A}=\frac{\epsilon^{3/4}}{\sqrt{(1-\beta
\epsilon)\mathrm{arccot}\left(\sqrt{\beta\epsilon}\right)+\sqrt{\beta\epsilon}}}.
\end{eqnarray}

\begin{figure}
\centering
\includegraphics[width=8cm]{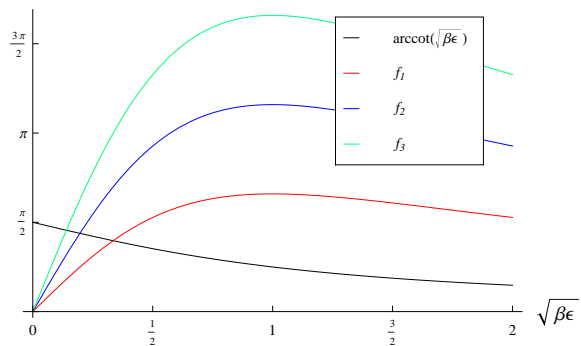}
\caption{\label{fig1}Schematic solutions of
$\mathrm{arccot}(x)=f_n(x)$, where $x=\sqrt{\beta\epsilon}$,
$f_n(x)=\frac{1+n\pi/(\alpha\sqrt{\beta})}{x+1/x}$, and
$\alpha=1/\sqrt{\beta}$.}
\end{figure}

By imposing the Hermicity condition (\ref{imc}) we find
\begin{eqnarray}\label{sin}
\sin\left[\alpha
\frac{1+\beta\epsilon}{\sqrt{\epsilon}}\mathrm{arccot}\left(\sqrt{\beta\epsilon}\right)-\alpha\sqrt{\beta}\right]=0,
\end{eqnarray}
so the quantization condition reads
\begin{eqnarray}\label{spec2}
\frac{1+\beta\epsilon}{\sqrt{\epsilon}}\mathrm{arccot}\left(\sqrt{\beta\epsilon}\right)-\sqrt{\beta}=\frac{n\pi}{\alpha},\hspace{1cm}n=1,2,\ldots\,.
\end{eqnarray}
It is straightforward to check that at the limit $\beta\rightarrow0$
the above condition agrees with the non-deformed energy condition,
i.e.,
\begin{eqnarray}\label{q-ordi}
\frac{\alpha}{2\sqrt{\epsilon}}=n,\hspace{2cm}n=1,2,\ldots\,.
\end{eqnarray}
Moreover, we have
\begin{eqnarray}\label{intP}
&&\hspace{-1cm}\int_{-\frac{1}{\sqrt{\beta}}}^{+\frac{1}{\sqrt{\beta}}}
\left(1-\beta p^2\right)\phi(p)\,\mathrm{d}
p=\frac{2\mathcal{A}}{\alpha}\sin\left[\alpha
\frac{1+\beta\epsilon}{\sqrt{\epsilon}}\mathrm{arccot}\left(\sqrt{\beta\epsilon}\right)-\alpha\sqrt{\beta}\right]=0.
\end{eqnarray}
The quantization condition (\ref{spec2}) can be rewritten as
\begin{eqnarray}\label{arccot}
\mathrm{arccot}(x_n)=\left(1+\frac{n\pi}{\alpha\sqrt{\beta}}\right)\left(x_n+\frac{1}{x_n}\right)^{-1},
\end{eqnarray}
where $x_n=\sqrt{\beta \epsilon_n}$. It is now obvious that for each
nonzero $n$ there always exists a solution for Eq.~(\ref{arccot}).
In figure \ref{fig1}, we have depicted the schematic solution for
the above equation. In the presence of just a minimal length, the
quantization condition is given by \cite{pedramH}
\begin{eqnarray}\label{minlength}
x_n=\frac{1}{2}\sqrt{1+\frac{2\alpha\sqrt{\beta}}{n}}-\frac{1}{2}.
\end{eqnarray}
In Table \ref{tab1}, we have reported the first ten solutions of the
quantization equations in ordinary quantum mechanics (\ref{q-ordi})
and two GUP scenarios (\ref{arccot},\ref{minlength}). As the table
shows because of the maximal momentum all the energy levels which
are proportional to $-x_n^2$ increase with respect to the presence
of just a minimal length.

\begin{table}
\caption{Solutions of the quantization equations in ordinary quantum
mechanics and two GUP scenarios for
$\alpha=1/\sqrt{\beta}$.}\label{tab1}
\begin{center}
\centering
\begin{tabular}{cccc}
$n$&&$x_n$&\\\hline
&\small{Absence} &&\small{Minimal Length}\\
&\small{of}&\small{Minimal Length}& \small{and} \\
&\small{GUP}&&\small{Maximal Momentum}\\\hline
1& $0.500000$& 0.366025  & 0.335027  \\
2& $0.250000 $& 0.207107  & 0.196343  \\
3& $0.166667 $& 0.145497  & 0.140027  \\
4& $0.125000$& 0.112372  & 0.109061  \\
5& $0.100000$& 0.091608  & 0.089388  \\
6& $0.083333$& 0.077350  & 0.075758  \\
7& $0.071429$& 0.066947  & 0.065749  \\
8& $0.062500$& 0.059017  & 0.058084  \\
9& $0.055556$& 0.052771  & 0.052023  \\
10&$0.050000$&0.047723  & 0.047110  \\\hline
\end{tabular}
\end{center}
\end{table}

To check the self-adjointness of the Hamiltonian, since the momentum
operator is obviously symmetric, we require that operator $1/X$ be a
symmetric operator on the set of eigenfunctions
\begin{eqnarray}
\left\langle \frac{1}{X}\varphi\bigg| \phi\right\rangle=\left\langle
\varphi\bigg| \frac{1}{X}\phi\right\rangle,
\end{eqnarray}
where $\phi$ and $\varphi$ belong to the domain of the Hamiltonian
and its adjoint, respectively. We can rewrite this condition using
the explicit expression for operator $1/X$ (\ref{X}) as
\begin{eqnarray}
\fl
i\int_{-\frac{1}{\sqrt{\beta}}}^{+\frac{1}{\sqrt{\beta}}}\left(1-\beta
p^2\right)\phi(p)\,\mathrm{d}
p\int_{-\frac{1}{\sqrt{\beta}}}^{p}\left(1-\beta
q^2\right)\varphi^*(q)\,\mathrm{d}
q+c^*\int_{-\frac{1}{\sqrt{\beta}}}^{+\frac{1}{\sqrt{\beta}}}\left(1-\beta
p^2\right)\phi(p)\,\mathrm{d} p\nonumber \\  \fl=-i
\int_{-\frac{1}{\sqrt{\beta}}}^{+\frac{1}{\sqrt{\beta}}}\left(1-\beta
p^2\right)\varphi^*(p)\,\mathrm{d}
p\int_{-\frac{1}{\sqrt{\beta}}}^{p}\left(1-\beta
q^2\right)\phi(q)\,\mathrm{d}
q+c\int_{-\frac{1}{\sqrt{\beta}}}^{+\frac{1}{\sqrt{\beta}}}\left(1-\beta
p^2\right)\varphi^*(p)\,\mathrm{d} p.\hspace{1cm}
\end{eqnarray}
Using the identity
\begin{eqnarray}
\fl\int_{-\frac{1}{\sqrt{\beta}}}^{+\frac{1}{\sqrt{\beta}}}f(p)\,\mathrm{d}
p\int_{-\frac{1}{\sqrt{\beta}}}^{p}g(q)\,\mathrm{d} q=
\int_{-\frac{1}{\sqrt{\beta}}}^{+\frac{1}{\sqrt{\beta}}}g(p)\,\mathrm{d}
p\bigg[\int_{-\frac{1}{\sqrt{\beta}}}^{+\frac{1}{\sqrt{\beta}}}f(q)\,\mathrm{d}
q-\int_{-\frac{1}{\sqrt{\beta}}}^{p}f(q)\,\mathrm{d} q\bigg],
\end{eqnarray}
we obtain
\begin{eqnarray}
&&\hspace{-2cm}i\int_{-\frac{1}{\sqrt{\beta}}}^{+\frac{1}{\sqrt{\beta}}}\left(1-\beta
p^2\right)\phi(p)\,\mathrm{d}
p\int_{-\frac{1}{\sqrt{\beta}}}^{+\frac{1}{\sqrt{\beta}}}\left(1-\beta
q^2\right)\varphi^*(q)\,\mathrm{d}
q\nonumber\\&&\hspace{-2cm}+c^*\int_{-\frac{1}{\sqrt{\beta}}}^{+\frac{1}{\sqrt{\beta}}}\left(1-\beta
p^2\right)\phi(p)\,\mathrm{d}
p-c\int_{-\frac{1}{\sqrt{\beta}}}^{+\frac{1}{\sqrt{\beta}}}\left(1-\beta
p^2\right)\varphi^*(p)\,\mathrm{d} p=0.\hspace{.7cm}
\end{eqnarray}
Therefore, because of (\ref{intP}) we have
\begin{eqnarray}\label{con-sym}
\int_{-\frac{1}{\sqrt{\beta}}}^{+\frac{1}{\sqrt{\beta}}}\left(1-\beta
p^2\right)\varphi^*(p)\,\mathrm{d} p=0.
\end{eqnarray}
These results show that the domains of $H$ and $H^\dagger$ coincide
and the Hamiltonian is rendered a true self-adjoint operator, i.e.,
$H=H^\dagger$ and
\begin{eqnarray}
\hspace{-1cm}{\cal D}(H)={\cal D}(H^{\dagger})=&&\bigg\{\phi\in{\cal
D}_{\mathrm{max}}\left(\frac{-1}{\sqrt{\beta}},\frac{+1}{\sqrt{\beta}}\right);\nonumber\\
&&\int_{-\frac{1}{\sqrt{\beta}}}^{+\frac{1}{\sqrt{\beta}}}\left(1-\beta
p^2\right)\phi(p)\,\mathrm{d} p=0\bigg\}.
\end{eqnarray}

\section{Coordinate space representation}
The eigenfunctions of the position operator satisfy the following
eigenvalue equation
\begin{eqnarray}
X\,u_x(p)=x\,u_x(p),
\end{eqnarray}
where $u_x(p)=\langle p|x\rangle$. In momentum space we have
\begin{eqnarray}
\frac{i}{1-\beta p^2}\partial_pu_x(p)=x\, u_x(p).
\end{eqnarray}
This equation can be solved to obtain the position eigenvectors
\begin{eqnarray}
u_x(p)=\mathcal{N} \exp\left[-i px
\left(1-\frac{\beta}{3}p^2\right)\right].
\end{eqnarray}
The eigenfunctions are normalizable
\begin{eqnarray}
1=\mathcal{N}\mathcal{N}^*\int_{-1/\sqrt{\beta}}^{+1/\sqrt{\beta}}\,\mathrm{d}
p \left(1-\beta
p^2\right)=\frac{4\mathcal{N}\mathcal{N}^*}{3\sqrt{\beta}},
\end{eqnarray}
and we finally obtain
\begin{eqnarray}
u_x(p)=\frac{\sqrt{3\sqrt{\beta}}}{2} \exp\left[-ip x
\left(1-\frac{\beta}{3}p^2\right)\right].
\end{eqnarray}
Now using the completeness relation
\begin{eqnarray}
\int_{-\frac{1}{\sqrt{\beta}}}^{+\frac{1}{\sqrt{\beta}}}\,\mathrm{d}
p\,\left(1-\beta p^2\right) |p\rangle\langle p|=1,
\end{eqnarray}
we find the wave function in coordinate space
\begin{eqnarray}\label{foo}
\psi(x)=\frac{\sqrt{3\sqrt{\beta}}}{2}
\int_{-\frac{1}{\sqrt{\beta}}}^{+\frac{1}{\sqrt{\beta}}}
\left(1-\beta p^2\right)e^{i
px\left(1-\frac{\beta}{3}p^2\right)}\phi(p)\,\mathrm{d} p.
\end{eqnarray}
Now Eq.~(\ref{intP}) implies
\begin{eqnarray}\label{psi}
\psi(x)\bigg|_{x=0}=0.
\end{eqnarray}
So the coordinate space wave functions vanish at the origin, i.e.,
they obey the Dirichlet boundary condition.

\section{Single-valuedness criteria}
Alternatively, one may use the requirement of single-valuedness of
eigenfunctions (\ref{sol}) to find the quantization condition which
leads to
\begin{eqnarray}\label{spec1}
\frac{\alpha\left(1+\beta\epsilon\right)}{2\sqrt{\epsilon}
}=m,\hspace{3cm}m=1,2,\ldots\,.
\end{eqnarray}
However, the eigenfunctions obeying quantization condition
(\ref{spec1}) do not satisfy (\ref{intP}) and (\ref{psi}) and the
Hermicity condition (\ref{imc}) therefore fails. Comparison between
the two quantization conditions (\ref{spec2}) and (\ref{spec1})
shows that
\begin{eqnarray}\label{sing}
m&=&\frac{n\pi+\alpha\sqrt{\beta}}{2\,\mathrm{arccot}\left(\sqrt{\beta\epsilon}\right)},\\
&=&n+\left(2n+\frac{\alpha}{\sqrt{\epsilon}}\right)\frac{\sqrt{\beta
\epsilon}}{\pi}+2\left(2n+\frac{\alpha}{\sqrt{\epsilon}}\right)\frac{\beta
\epsilon}{\pi^2}+\cdots.\hspace{.7cm}
\end{eqnarray}
So the single-valuedness criteria of the eigenfunctions ($m\in$
integers) is only valid at the limit $\beta\rightarrow0$, i.e., the
absence of GUP. However, since for $\beta=0$ all eigenfunctions
satisfy
\begin{eqnarray}
\int_{-\infty}^{+\infty} \phi(p)\,\,\mathrm{d} p=0=\psi(0),
\end{eqnarray}
regardless of their energy, the single-valuedness condition is still
valid at this limit.

Note that in ordinary quantum mechanics, the wave function
associated with the particle must be single valued of its argument.
Because, when two values are found, it means that the particle
exists in two different places which is impossible when we consider
particles as point-like objects. However, in the presence of the
minimal uncertainty relation, the assumption of the point-like
particles is no longer valid and the failure of the
single-valuedness criteria is not problematic in this framework due
to the fuzziness of space. Moreover, as Eq.~(\ref{sing}) shows, this
condition is only slightly broken in practise (small values of
$\beta$) which agrees with the smallness of the uncertainty in
position measurement (\ref{minl}). It is worth mentioning that the
failure of the single-valuedness condition is well-known in the
framework of the generalized uncertainty principle. For instance, as
it is shown in the seminal paper by Kempf \emph{et al}., even the
first GUP solutions such as the eigenfunctions of the position
operator and the maximal localization states are not single valued
in the presence of just the minimal length \cite{7}.

\section{Semiclassical solutions}
The semiclassical energy spectrum is given by the Bohr-Sommerfeld
quantization condition
\begin{eqnarray}
\oint p\,\,\mathrm{d} x=2n\pi,\hspace{1.5cm}n=1,2,\ldots\,.
\end{eqnarray}
The corresponding classical Hamiltonian to the hydrogen atom problem
is
\begin{eqnarray}\label{eqP0}
H(x,p)=p^2-\frac{\alpha\left(1-\beta p^2\right)}{x},
\end{eqnarray}
where we used
\begin{eqnarray}
X=\frac{1}{1-\beta p^2}x,\hspace{1.5cm}x=i
\frac{\,\mathrm{d}}{\,\mathrm{d} p},
\end{eqnarray}
and neglected the ordering problem in classical domain. Since the
Hamiltonian is conserved, i.e. $H(x,p)=E=-\epsilon$, we can express
$p$ as a function of $x$, namely,
\begin{eqnarray}
p=\left(\frac{\alpha-\epsilon\, x}{x+\alpha\beta}\right)^{1/2}.
\end{eqnarray}
When the particle leaves the origin in positive direction, $x$
changes from $0$ to ${\alpha}/{\epsilon}$. So $\displaystyle\oint
p\,\,\mathrm{d} x=2\int_{0}^{\frac{\alpha}{\epsilon}}p\,\,\mathrm{d}
x$ and for the negative energy bound states we find
\begin{eqnarray}
2n\pi&=&2\int_{0}^{\frac{\alpha}{\epsilon}}\left(\frac{\alpha-\epsilon\,
x}{x+\alpha\beta}\right)^{1/2}\,\,\mathrm{d}
x\nonumber\\
&=&\frac{2\alpha\left(1+\beta\epsilon\right)}{\sqrt{\epsilon}}\mathrm{arccot}\left(\sqrt{\beta\epsilon}\right)-2\alpha\sqrt{\beta},
\end{eqnarray}
which exactly agrees with Eq.~(\ref{spec2}). The validity of the
semiclassical approximation for this modified quantum mechanics is
also discussed in \cite{pedramHigh2}.

\section{Conclusions}
In this paper, we have considered the problem of the one-dimensional
hydrogen atom in the presence of both a minimal length uncertainty
and a maximal momentum and found exact energy eigenvalues and
eigenfunctions. By imposing the Hermicity condition on the
Hamiltonian, the quantization condition is uniquely determined and
the Hamiltonian is rendered self-adjoint. We showed that the
single-valuedness condition is only valid  for the zero deformation
parameter and the coordinate space wave functions vanish at the
origin. Moreover, similar to the case where there is just a minimal
length \cite{pedramH}, the semiclassical energy spectrum exactly
coincides with the quantum mechanical results.


\section*{References}
\begin{thebibliography}{99}
\bibitem{felder}    S. Hossenfelder, arXiv:1203.6191.
\bibitem{epl1}      P.~Pedram, Europhys. Lett. \textbf{89} (2010) 50008.
\bibitem{epl2}      P.~Pedram and K. Nozari K. Europhys. Lett. \textbf{92} (2010) 50013.
\bibitem{jhep}      P.~Pedram, K. Nozari, and S.H. Taheri JHEP \textbf{1103} (2011) 093.
\bibitem{1}         G. Veneziano, Europhys. Lett. \textbf{2} (1986) 199.
\bibitem{2}         E. Witten, Phys. Today \textbf{49}  (1996) 24.
\bibitem{3-1}       D. Amati, M. Ciafaloni, G. Veneziano, Phys. Lett. B \textbf{216}  (1989) 41.
\bibitem{3-2}       D. Amati, M. Ciafaloni, G. Veneziano, Nucl. Phys. B \textbf{347}  (1990) 550.
\bibitem{3-3}       D. Amati, M. Ciafaloni, G. Veneziano, Nucl. Phys. B \textbf{403}  (1993) 707.
\bibitem{4}         K. Konishi, G. Paffuti, P. Provero, Phys. Lett. B \textbf{234}  (1990) 276.
\bibitem{5}         L.J. Garay, Int. J. Mod. Phys. A \textbf{10}  (1995) 145.
\bibitem{6}         M. Maggiore, Phys. Lett. B \textbf{319}  (1993) 83.
\bibitem{7}         A. Kempf, G. Mangano, R.B. Mann, Phys. Rev. D \textbf{52}  (1995) 1108.
\bibitem{8}         A. Kempf, G. Mangano, Phys. Rev. D \textbf{55}   (1997) 7909.
\bibitem{21}        J. Magueijo and L. Smolin, Phys. Rev. Lett. \textbf{88}  (2002) 190403, arXiv:hep-th/0112090.
\bibitem{22}        J. Magueijo and L. Smolin, Phys. Rev. D \textbf{71}  (2005) 026010, arXiv:hep-th/0401087.
\bibitem{23}        J.L. Cortes and J. Gamboa, Phys. Rev. D \textbf{71}  (2005) 065015, arXiv:hep-th/0405285.
\bibitem{11}        J.A. Reyes  and M. del Castillo-Mussot, J. Phys. A \textbf{32}  (1999) 2017.
\bibitem{12}        Y. Ran, L. Xue, S. Hu and R-K. Su, J. Phys. A \textbf{33}  (2000) 9265.
\bibitem{13}        A.N. Gordeyev and S.C. Chhajlany, J. Phys. A \textbf{30}  (1997) 6893.
\bibitem{14}        I. Tsutsui, T. Fulop and T. Cheon,  J. Phys. A \textbf{36}  (2003) 275.
\bibitem{15}        H.N. Nunez Yepez, C.A. Vargas and A.L.S. Brito,  Eur. J. Phys. \textbf{8}  (1987) 189.
\bibitem{Akhoury}   R. Akhoury and Y.-P. Yao, Phys. Lett. B \textbf{572}  (2003) 37.
\bibitem{Bouaziz}   D. Bouaziz, N. Ferkous, Phys. Rev. A \textbf{82}  (2010) 022105.
\bibitem{Brau}      F.~Brau, J.~Phys.~A \textbf{32}  (1999) 7691.
\bibitem{Benczik}   S. Benczik, L.N. Chang, D. Minic and T. Takeuchi, Phys. Rev. A \textbf{72}  (2005) 012104.
\bibitem{Tkachuk}   T.V. Fityo, I.O. Vakarchuk and V.M. Tkachuk, J. Phys. A  \textbf{39}   (2006) 2143.
\bibitem{pedramPRD} P.~Pedram, Phys.~Rev.~D \textbf{85}  (2012) 024016, arXiv:1112.2327.
\bibitem{pedramPLB} P.~Pedram, Phys.~Lett.~B \textbf{710}  (2012) 478.
\bibitem{pedramH}   P.~Pedram, J. Phys. A \textbf{45} (2012) 505304, arXiv:1203.5478.
\bibitem{pedramHigh}P.~Pedram, Phys. Lett. B \textbf{714}  (2012) 317, arXiv:1110.2999.
\bibitem{pedramHigh2}  P.~Pedram, Phys. Lett. B \textbf{718}  (2012) 638, arXiv:1210.5334.
\end{thebibliography}
\end{document}